\documentclass[authoryear,preprint,review,12pt]{elsarticle}

\usepackage{amssymb}
\usepackage{booktabs}
\usepackage[table,xcdraw]{xcolor}
\usepackage{graphicx}
\usepackage{svg}
\usepackage{float}
\usepackage{amsmath, mathtools}
\usepackage{mathrsfs}
\usepackage{gensymb}
\usepackage{multicol}
\usepackage{multirow}
\usepackage{tabularx}
\usepackage{geometry}
\usepackage{hyperref}
\usepackage{soul}
\usepackage{lineno}

\journal{arXiv}

\begin{document}

\begin{frontmatter}

\title{A Method for Classifying Snow Using Ski-Mounted Strain Sensors}

\author[inst1]{Florian McLelland}
\author[inst1]{Floris van Breugel}

\affiliation[inst1]{organization={Department of Mechanical Engineering},
            addressline={1664 N Virginia St}, 
            city={Reno},
            postcode={89557}, 
            state={NV},
            country={USA}}

\begin{abstract}
Understanding the structure, quantity, and type of snow in mountain landscapes is crucial for assessing avalanche safety, interpreting satellite imagery, building accurate hydrology models, and choosing the right pair of skis for your weekend trip. Currently, such characteristics of snowpack are measured using a combination of remote satellite imagery, weather stations, and laborious point measurements and descriptions provided by local forecasters, guides, and backcountry users. Here, we explore how characteristics of the top layer of snowpack could be estimated while skiing using strain sensors mounted to the top surface of an alpine ski. 
We show that with two strain gauges and an inertial measurement unit it is feasible to correctly assign one of three qualitative labels (powder, slushy, or icy/groomed snow) to each 10 second segment of a trajectory with 97\% accuracy, independent of skiing style. Our algorithm uses a combination of a data-driven linear model of the ski-snow interaction, dimensionality reduction, and a Na\"{\i}ve Bayes classifier. 
Comparisons of classifier performance between strain gauges suggest that the optimal placement of strain gauges is halfway between the binding and the tip/tail of the ski, in the cambered section just before the point where the unweighted ski would touch the snow surface. 
The ability to classify snow, potentially in real-time, using skis opens the door to applications that range from citizen science efforts to map snow surface characteristics in the backcountry, and develop skis with automated stiffness tuning based on the snow type. 
\end{abstract}

\begin{graphicalabstract}
\includegraphics{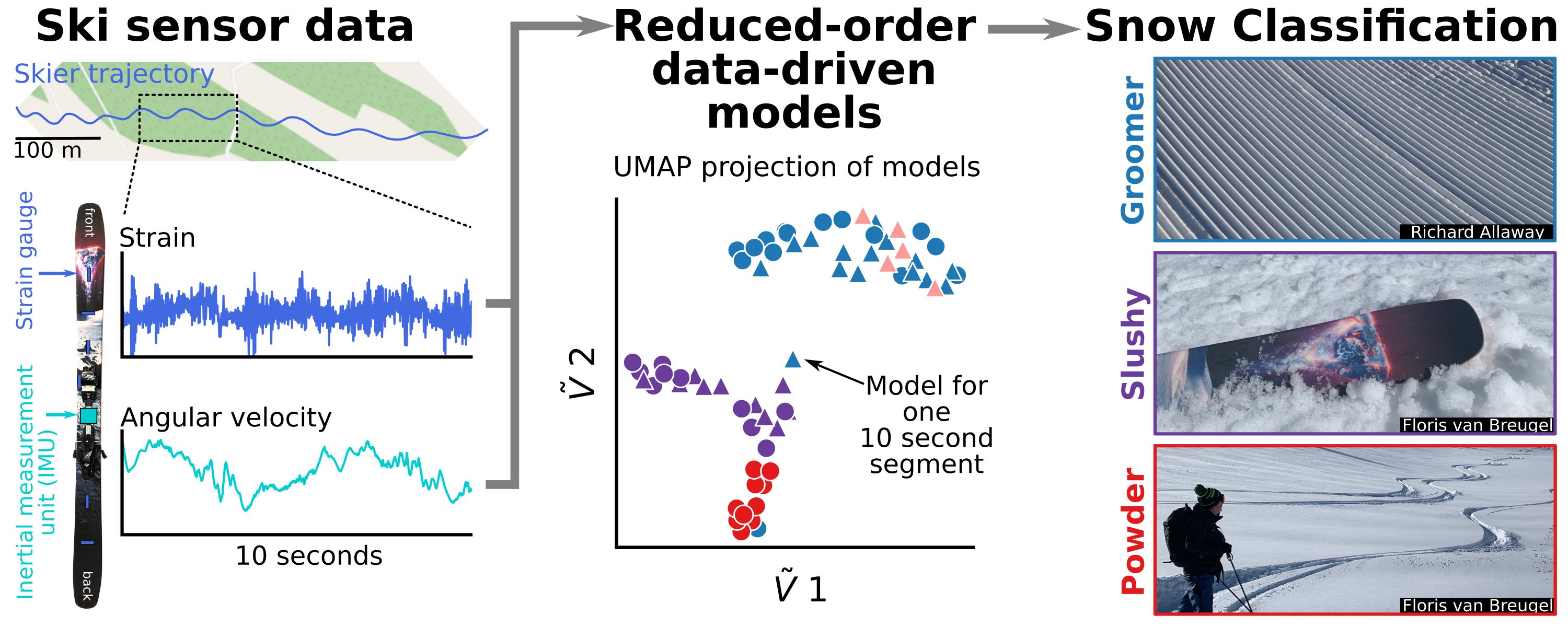}
\end{graphicalabstract}

\begin{highlights}
\item Ski strain can be used to classify snow as powder, slushy corn, or hard-packed snow 
\item Classification requires accounting for skier supplied torques, or a proxy thereof
\item Strain sensors should be placed midway between the binding and tip and/or tail 
\item Time series data from four consecutive turns is sufficient for classification
\end{highlights}


\end{frontmatter}



\section{Introduction}
\label{sec:introduction}
Accurate measurements of snow characteristics, such as depth, density, and snow water equivalent, are necessary for myriad applications including hydrology models, avalanche predictions, and gear and terrain selection for snow sports such as skiing and snowboarding. Hydrology models in alpine regions and downstream basins require accurate estimates of both depth and snow water equivalent to predict the amount of water that will become available for ecosystems and human activities over the course of the year \citep{kane1991snow,liston1998snow}. Non-uniform topography and vegetation across landscapes results in spatial variability of snow depth, density, and snow water equivalent \citep{liston1998snow}. Understanding this variability has been shown to be important for constructing accurate hydrology models \citep{liston1999interrelationships}. The spatial variability of density also helps to determine its stability, which is critical to understand from a snow safety perspective. Each year, many people are injured or killed by avalanches, some of which could be prevented by improved snow-pack stability models \citep{morin2020application}. Finally, ensuring an enjoyable experience on skis or a snowboard can benefit from an understanding of how snow density varies. Snow density starts out very low with fresh powder snow, but as the snow melts and compacts it becomes more dense, eventually forming ice. The density of snow can range from 430 to 660 $\mathrm{kg}/\mathrm{m}^3$ \citep{federolf2006deformation}, which affects a skiers ability to properly execute a turn \citep{mossner2009science}. Elegantly skiing over various densities of snow requires skill, but can also be improved by using a ski with the correct stiffness for the specific terrain and snow characteristics \citep{nachbauer2004effects, foss2007reducing, lind2013physics}. Unfortunately, despite the value of spatial information about density and snow water equivalent, currently available sensing technologies are limited in their abilities to provide this data. 

There are a wide variety of currently available snow sensing methods including both remote satellite approaches, and ground based measurements. Synthetic Aperture Radar (SAR) is a satellite based sensor that can provide accurate snow cover extent (SCE), grain size and snow water equivalent (SWE), even with polar darkness and cloud obstructions \citep{tsai2019combination}. SAR, however, struggles to provide accurate SWE measurements of tree covered regions \citep{tsai2019remote}. Active Microwave remote sensing is another remote sensing method, but is negatively effected by back-scattering from steep topographic geometry and vegetation \citep{dozier2016estimating}. To augment remote sensing methods, a number of ground based methods can be used. Some methods include snow pits where an expert digs a vertical hole in the snow-pack and analyzes the layers of snow, and ultrasonic probes to measure snow depth \citep{ryan2008evaluation,eisen2008ground}. All available ground based methods, however, provide limited spatial information because they require either stationary weather stations, or a substantial amount of manual labor that must be performed by trained experts.

The goal of this paper is to provide a proof of concept for a new ground based sensing method that would allow non-experts to contribute spatial measurements of snow characteristics related to density and snow water equivalent to help improve models of snow pack. Any experienced skier or snowboarder will have an intuitive sense of whether they are skiing on fresh powder, a groomed ski run at a resort, hard packed ice, slushy spring corn snow, or more complex conditions. This intuitive sense is largely driven by the vibrations that can be felt through the ski while making left and right turns. We set out to develop a data processing algorithm that could automate this process given strain and angular velocity measurements from a ski, making it possible for any skier to measure location specific qualitative snow characteristics simply by skiing through the terrain. Our algorithm uses a control theoretic framework to build data-driven models of the ski-snow interactions every four turns, and then uses a classifier to assign a qualitative snow type to each model. Our approach is able to accurately classify powder, groomer, and hard-packed (groomer and icy) snow with roughly 97\% accuracy given sensor data from just four turns. Qualitative snow characteristics such as powder, slushy, and hard-packed snow can be related to the snow grain size, which in turn has a roughly linear relationship to snow water equivalence when looking in the range of 0.4 to 0.75 albedo \citep{nolin1993estimating, dozier1981effect}. Thus, our approach demonstrates the feasibility of developing a citizen science platform for collecting spatially detailed snow characteristics to augment existing remote and ground based sensing methods, which could help to improve models of snow pack. 

\section{Methods}

\subsection{Hardware and data collection}

To collect data to use in developing a snow classification algorithm, we equipped a ski with an array of sensors and recruited a skier to ski on four qualitatively different types of snow. The following subsections provide details on the skis, snow, terrain, sensors, and data collection.

\subsubsection{Skis, Snow, Terrain and Skier Trajectories}

We used a pair of Wildcat skis (length 179 cm, tip and tail width 141mm, waist 116 mm) provided by Moment Skis, Reno, NV. The right foot ski was outfitted with an array of sensors, which were connected to microcontrollers for data logging (see next section). For consistency, we recruited one skier whose height is 6’ 1” and weight is 155lbs. The skier skied on various snow conditions at the Mount Rose Ski Resort near Reno, NV. 

All ski runs were done on either the Big Bonanza or Silver Dollar run, which offer consistent terrain and aspect, free of moguls, bumps, and jumps. The consistent aspect helped to ensure that snow characteristics throughout the length of each run were relatively uniform. We assigned one of four qualitative classes of snow to each ski run: powder, slushy, groomer, and icy snow. Powder snow is soft pillow-like and very light. Slushy snow (i.e. spring corn snow) is also soft, however it is much heavier and wetter. Icy and groomer snow types are both hard and compact, the main difference being that in icy snow it is more difficult to maintain consistent ski edge contact, resulting in more lateral ski movement. Each snow type was sampled on a separate day and was not sampled the same amount due to limited access to powder and icy snow. 

\begin{figure}[H]
	\includegraphics[scale=0.9]{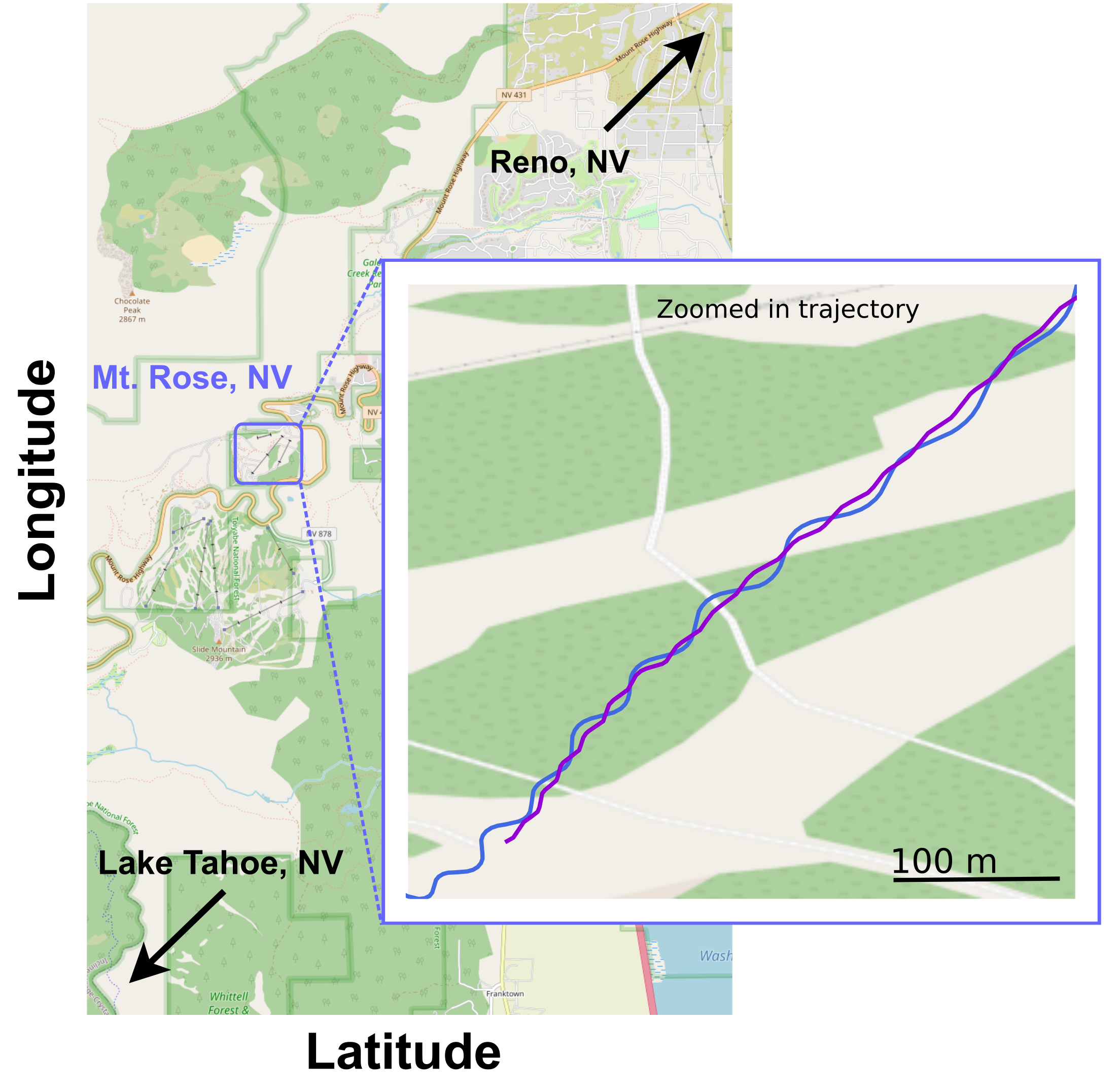}
	\centering
	\caption{\footnotesize Example ski trajectories at Mt. Rose Ski Resort, NV. Inset shows two characteristics trajectories, one with big and slow turns (blue) and one with tight higher frequency turns (purple). Map graphics from OpenStreetMap.}
	\label{fig:trajectories}
\end{figure}

To determine the impact of skiing style on our classification, the skier executed two types of ski trajectories: either big slow turns, or tight fast turns (Fig. \ref{fig:trajectories}. Table 2 shows the details of each ski run and quantity of ski runs per snow type.

\begin{table}[H]
\footnotesize
\centering
\caption{\label{tab:experiments} \footnotesize Summary of data collected with sensor equipped skis.}
\begin{tabular}{p{1cm} p{0.75cm} p{1.5cm} p{1cm} c}
\\\hline
\multicolumn{5}{c}{Location: Mt. Rose, NV}\\\hline
\textbf{Snow Type} & \textbf{Temp F\degree} & \textbf{Turning Style} & \textbf{Date} & \textbf{Notes} \\\hline
\multirow{2}{1cm}{Groomer} & \multirow{2}{0.75cm}{25} & Big x3 & \multirow{2}*{01/26/21} & \multirow{2}{2.5cm}{Smooth groomer}\\
& & Tight x3 & &\\
\hline
\multirow{2}{1cm}{Powder} & \multirow{2}{0.75cm}{20} & Big & \multirow{2}*{01/28/21} & One run of pristine\\
& & Big & & powder, one \\ & & & &  partially skied out\\\hline
\multirow{2}{1cm}{Icy} & \multirow{2}{0.75cm}{20} & Tight & \multirow{2}*{03/30/21} & \multirow{2}{2.5cm}{Groomed, but hard and icy}\\
& & Big & &\\\hline
\multirow{3}{1cm}{Slushy} & \multirow{3}{0.75cm}{40} & Tight x2 & \multirow{3}*{04/01/21} & \multirow{3}{2.5cm}{Some 6" piles of slush from previous skiers}\\
& & Big x2 & &\\
\\\hline

\end{tabular}
\end{table}

\subsubsection{Sensors and Ski Setup}

\begin{figure}[H]
	\includegraphics[scale=0.9]{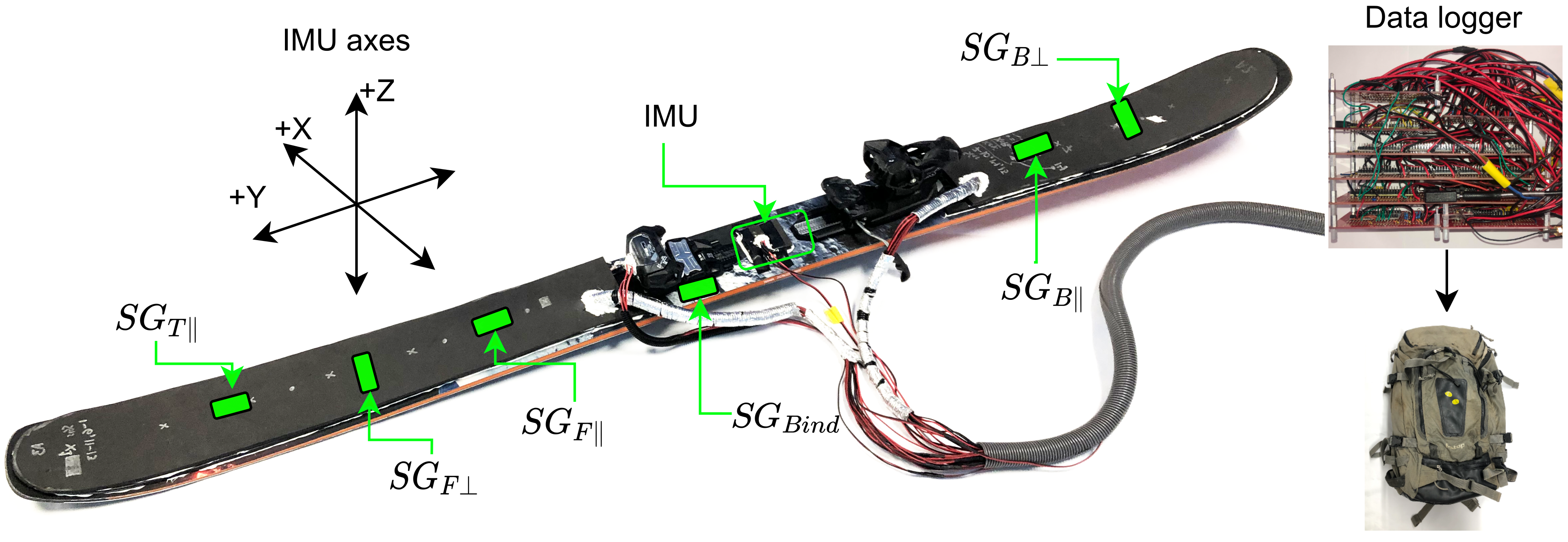}
	\centering
	\caption{\footnotesize Moment Wildcat 108 ski used in our experiments, outfitted with strain sensors on the front and back of the ski depicted in green. An IMU is located between the toe and heel binding points. All sensors are covered by closed cell foam padding, and wires go to a data logger stored in a backpack. }
	\label{fig:ski_setup}
\end{figure}

Inspired by the prevalence of strain sensors found on flexible insect wings \citep{dickerson2014control, mohrenpnas}, and recent success in using wing-mounted strain gauges on an aircraft to detect gusts of wind \citep{windsorstrain}, we chose to use strain gauges to measure the bending and vibrations of the ski. We mounted a total of six strain gauges to the top of the ski: four were oriented parallel to the length of the ski, and two perpendicular to the length of the ski (Fig. \ref{fig:ski_setup}). One of the parallel strain gauges was placed directly next to the binding. Strain gauges were configured in a quarter Wheatstone bridge circuit using 120$\Omega$ resistors to match the strain gauge resistance. The output voltage from the Wheatstone circuit was connected to a 24-bit 1KHz analog to digital converter (ADC) by Protocentral, which sent the digital data to Teensy 3.5 microcontrollers via Serial Peripheral Interface (SPI).

To record a proxy for the forces and torques applied to the ski, we mounted an inertial measurement unit (IMU) to the ski below the ski boot. The IMU provided measurements of angular velocity and linear acceleration at a rate of 200 Hz, which was transmitted to a separate Teensy via the Inter-Integrated Circuit (I2C) Protocol for serial communication. 

To examine the shape and velocity of the skier trajectories, we used a GPS system with an antenna mounted to the skier's helmet. The GPS sent its data via serial communication to the same Teensy connected to the IMU, but at a rate of 10Hz. The GPS data was not used as a part of our classification algorithm. All sensors and sensor specifications are tabulated in Table \ref{tab:sensors}.

\begin{table}[H]
\footnotesize
\centering
\caption{\label{tab:sensors} \footnotesize Sensors used on the ski in experiments include strain gauges, an IMU, and a GPS. All three types of sensors sample at different rates requiring signal interpolation.}
\begin{tabular}{@{}cccc@{}}
\\
\toprule
\textbf{Sensors} & \textbf{Part Number} & \textbf{Sample Rate} & \textbf{Company} \\ \midrule
\begin{tabular}[c]{@{}c@{}}Strain gauge \\ (120 $\Omega$),\\ ADC (24 bit)\end{tabular} & \begin{tabular}[c]{@{}c@{}}RS pro: \\ 632180, \\ADS1220 \end{tabular} & 1KHz & \begin{tabular}[c]{@{}c@{}}Allied,\\ Protocentral\end{tabular} \\ \midrule
IMU (6Dof) & LMS9DS1 & 200Hz & SparkFun \\ \midrule
GPS & ZOE-M8Q & 10Hz & SparkFun \\ \bottomrule
\end{tabular}
\end{table}

Only the right ski was equipped with sensors, which were covered by closed cell foam to provide impact protection, minimize water entry points, and reduce temperature fluctuations (which can impact strain gauge measurements). We applied a silicone coating to each strain gauge and the IMU to waterproof the circuitry. Wires from each sensor were routed inside the skier’s pant leg into a backpack where the data acquisition unit was housed.
    
\subsubsection{Data Logging}

Each of our sensors provided data at different sampling rates, and we used a total of four Teensy microcontrollers to record the data. To align the data, each Teensy was programmed to receive a random pulse signal from the ``master" Teensy collecting IMU and GPS data. We used this random pulse to manually align the signals to have a consistent timestamp. Once the signals were aligned, the IMU and GPS signals were linearly interpolated to the time step of the strain gauge. All data was collected on micro-SD cards in binary form. The teensies and ADC converters were housed in a plastic box within the skier’s backpack and powered by 5V battery packs. Analysis was done in post on a separate computer using custom software written in Python.

\begin{figure}[H]
    \centering
    \includegraphics[scale=0.85]{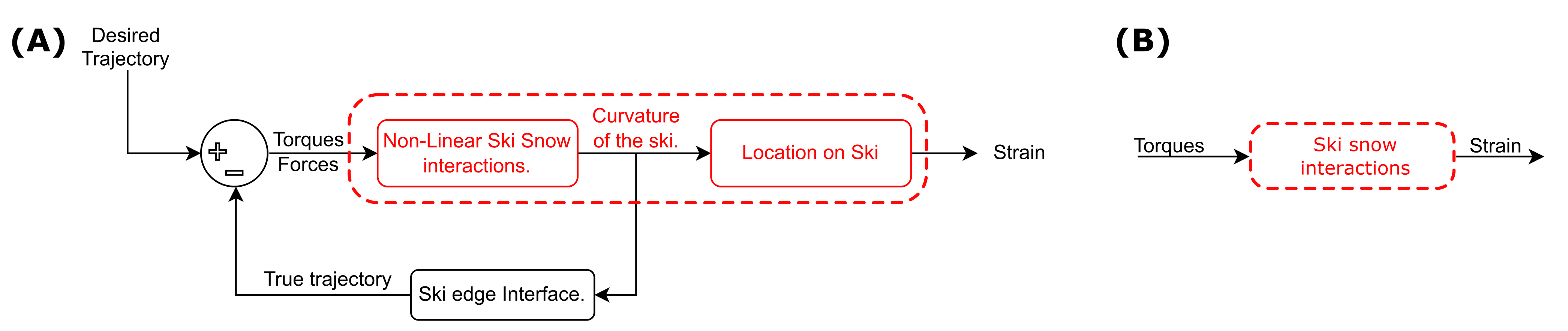}
    \caption{\footnotesize Skier-ski-snow block diagram. (A) The skier applies torques to the ski to navigate the terrain according to a desired trajectory. The ski-snow dynamics result in deformations in the ski, causing strain that can be measured by strain sensors sensors. (B) Simplified block diagram that we use throughout the paper.}
    \label{fig:block_diagram_complete}
\end{figure}

\subsection{Modeling} \label{Modeling}

To characterize snow type given IMU and strain sensor measurements from the ski we began by building a collection of linear data-driven models of the ski-snow dynamics associated with each 10 second segment. We then trained a classifier to assign one of four qualitative snow type labels to each model collection. This section provides a detailed description of our modeling framework.   

The interactions between an alpine ski and snow are extremely complex due to myriad non-linearities. The four dominant ski snow interactions are: smooth flat snow contact, self-excited vibration due to lateral sliding which occurs on harder snow, free vibrations when the ski is not in contact with the snow, and pure edge contact without sliding \citep{lind2013physics, foss2007reducing, gosselin2021effect}. The non-continuous nature of these four modes is what creates a non-linear system. Given the complexities of constructing non-linear data-driven models, we opted to simplify the problem and treat it as a linear system. We found that this simplification works well for our dataset, however, future efforts may require considering more of these non-linearities to provide more robust and detailed snow classification.

\subsubsection{Block diagram and model}

The primary measurements we used for our snow classification were strain measurements made at several locations on the ski. The ski's strain is determined by both the ski-snow interactions and the torques and forces applied by the skier to the ski. To understand this relationship, we begin by modeling the skier, ski, and snow using a simple feedback control diagram (Fig. \ref{fig:block_diagram_complete}A). In this model, the skier uses visual and proprioceptive feedback to compare their true trajectory to a mental image of their desired trajectory, and applies torques and forces to the ski to make the necessary corrections \citep{muller1994analysis}. These forces and torques act as inputs to the ski-snow dynamics. Together, these inputs and ski-snow dynamics determine the time varying behavior of the ski's curvature. This curvature determines both the ski's trajectory, and strain at each location on the ski. 

In traditional alpine skiing, the majority of the skier’s weight is applied to the outside ski and a slight upward force is applied to the inside ski causing that ski to have less contact with the snow. Although forces are applied by the skier, torque is the dominant skier input because turning is primarily caused by parabolic edge-snow contact which is accomplished by torques applied by the skier \citep{lind2013physics}. Thus, we can build a simplified linear dynamical model of the ski-snow interactions by comparing time series data of strain from each strain gauge to the 3-dimensional torques (\ref{fig:block_diagram_complete}B). Because the ski and location of each strain sensor remained constant across all of our experiments, differences between the resulting models can be used to characterize the snow.

Using linear systems theory we can construct single-input single-output models in the frequency domain, referred to as ``transfer functions". Na\"{i}vely, one could construct a data-driven transfer function for each strain gauge (e.g. $SG_{F||}$) and torque about each axis (e.g. $\tau_x$) by calculating the magnitude and phase that describe the complex ratio of their Discrete Fourier Transforms (DFTs). For example, for strain gauge $SG_{F||}$ we could construct three models:
\begin{equation}
\begin{split}
    \text{ski-snow interactions for } SG_{F||} = \\
    \bigg[ \frac{DFT(SG_{F||})}{DFT(\tau_x)}; \frac{DFT(SG_{F||})}{DFT(\tau_y)}; \frac{DFT(SG_{F||})}{DFT(\tau_z)} \bigg].
\end{split}
\end{equation}
In practice there are three substantial challenges with this approach: (1) the noise in our strain gauge measurements makes reliably estimating phase impossible, (2) torque is difficult to measure, or estimate, without introducing several noise amplifying steps, and (3) calculating a ratio of two Discrete Fourier Transforms of real world data can lead to non-physical results when the magnitude of either the numerator or denominator approaches zero. To address the first challenge we focused our modeling efforts exclusively on the magnitude of the ratio of the output and input, and omit the phase relationship from our analysis. The following two subsections detail our solutions for mitigating the second and third challenges.

\subsubsection{Angular velocity as a proxy for torque}

Measuring the true skier input of torque in all three axes is challenging.
However, by modeling the ski and boot as a rigid body it is possible to estimate the torque given measurements from the IMU and a rigid body model of the ski and boot. The moment on the ski can be defined using the following equation,

\begin{equation}
\footnotesize
\label{eq:torque_eqn}
\begin{split}
\prescript{N}{}{\Vec{M}}^{B/B_p} \;\;\overset{\Delta}{=} \;\;&\Vec{\Vec{I}}^{B/B_p} \cdot \prescript{N}{}{\Vec{\dot{\omega}}}^B \\
+ \;\;&\prescript{N}{}{\Vec{\omega}}^B \times (\Vec{\Vec{I}}^{B/B_p} \cdot \prescript{N}{}{\Vec{\omega}}^B) \\
+ \;\;&\Vec{r}^{B_cm/B_p} \times m^B \prescript{N}{}{\Vec{a}}^B_p,
\end{split}    
\end{equation} 
where each terms is defined as:
\begin{enumerate}
    \item[] $\prescript{N}{}{\Vec{M}}^{B/B_p}$ represents the three dimensional torque applied to body $B$ at point $B_p$ in reference frame $N$;
    \item[] $\Vec{\Vec{I}}^{B/B_p}$ represents the inertia dyadic of body $B$ at point $B_p$;
    \item[] $\prescript{N}{}{\Vec{\dot{\omega}}}^B$ represents the three dimensional angular acceleration of body $B$ in reference frame $N$;
    \item[] $\prescript{N}{}{\Vec{\omega}}^B$ represents the three dimensional angular velocity of body $B$ in reference frame $N$;
    \item[] $\Vec{r}^{B_cm/B_p}$ represents the three dimensional distance of point $B_p$ to the center of gravity of body $B$;
    \item[] $m^B$ represents the mass of body $B$;
    \item[] $\prescript{N}{}{\Vec{a}}^B_p$ represents the three dimensional linear acceleration of point $B_p$ in reference frame $N$. 
\end{enumerate}

The IMU provides measurements of linear acceleration $(a)$ and angular velocity $(\omega)$ for each of the three axes. We estimated the angular acceleration $(\dot{\omega})$ by numerically differentiating the angular velocity measurements provided by the IMU. To get smooth estimates of $\dot{\omega}$ we applied a third-order Butterworth smoother to $\omega$ \citep{Butterworth1930} with a cutoff frequency of 100 Hz, followed by a finite difference calculation, using the implementation provided by the Pynumdiff python package \citep{van2022pynumdiff}.

\begin{figure}[H]
    \centering
    \includegraphics[scale=0.9]{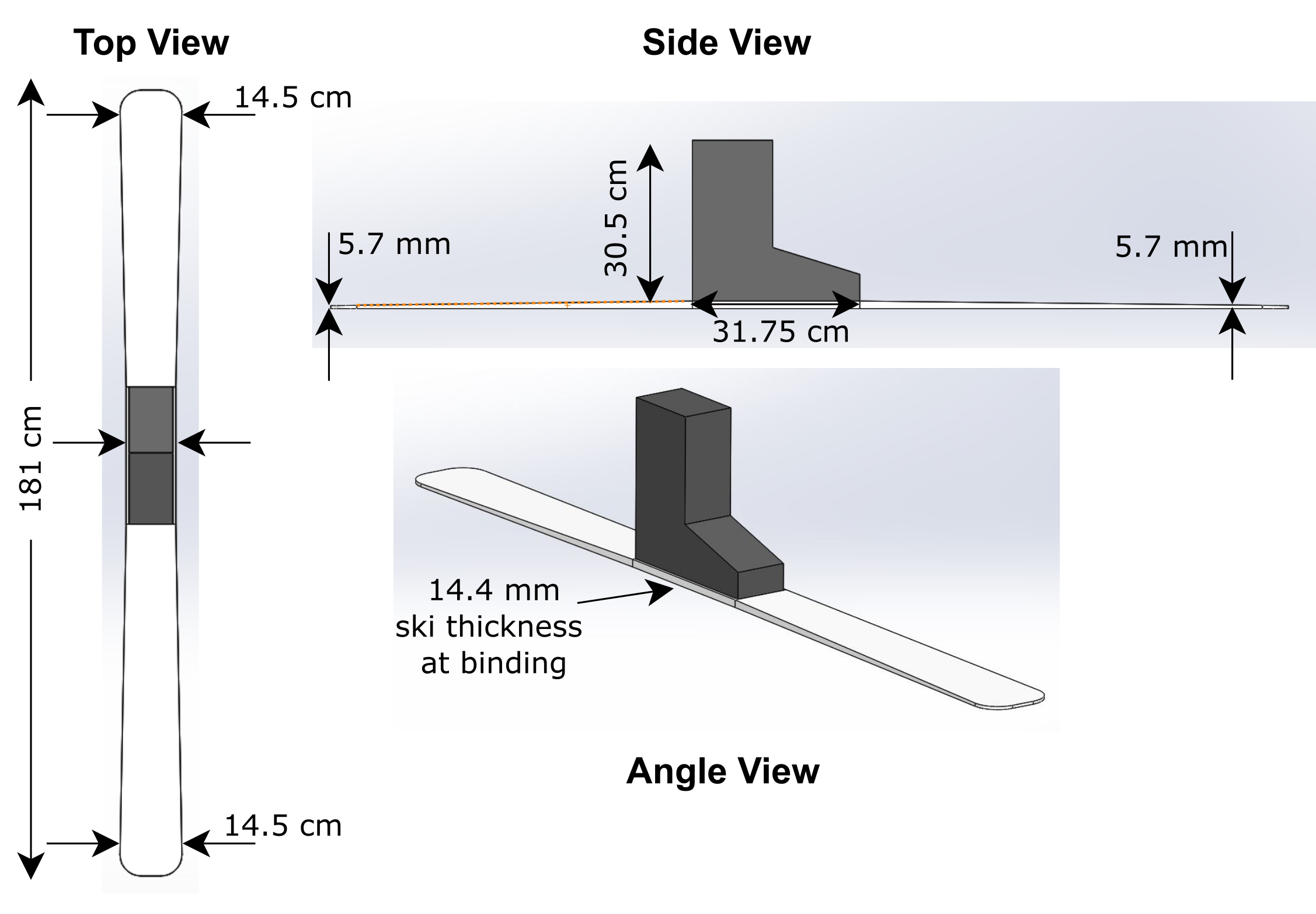}
    \caption{\footnotesize Solidworks model of the ski and boot, based on dimensions for the Wildcat ski used for our experiments. We used this model to determine the inertia dyadic 
(Table \ref{tab:inertia_dyadic}) used to estimate torque.}
    \label{fig:solidworks_model}
\end{figure}

We estimated $I$ and $m$ by building a simplified physical model of the ski and boot in SolidWorks 2020 computer-aided design software (Fig. \ref{fig:solidworks_model}). We assumed uniform density through the ski and boot, and chose ABS plastic as the material. The material choice is arbitrary as it only scales the magnitude of the inertia dyadic. Table \ref{tab:inertia_dyadic} displays the resulting moments and products of inertia of the ski boot and ski. 

\begin{table}[H]
\centering
\caption{\label{tab:inertia_dyadic} \footnotesize Moments and products of the rigid ski-boot combination, displayed in units of $\mathrm{kg}/\mathrm{m}^2$.}
\begin{tabular}{l|lll}
 & $I_x$ & $I_y$ & $I_z$ \\\hline
$I_x$ & 0.77 & 0.01 & 0.00 \\
$I_y$ & 0.01 & 0.16 & -0.06 \\
$I_z$ & 0.00 & -0.06 & 0.62
\end{tabular}
\end{table}

\begin{figure}[H]
    \centering
    \includegraphics[scale=1.0]{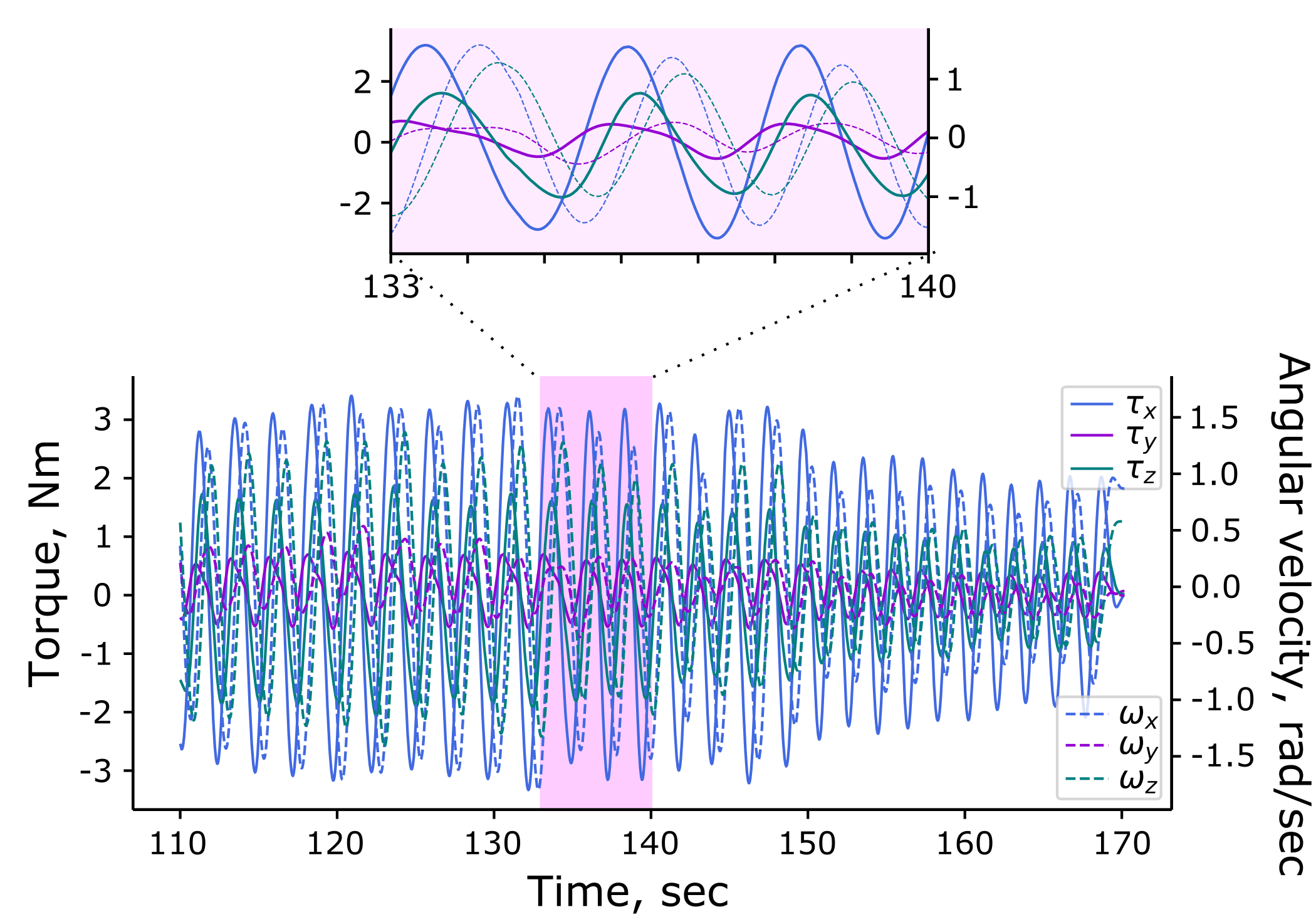}
    \caption{\footnotesize Smoothed torques (solid) plotted over smoothed angular velocities (dashed) for one ski run. The measured angular velocities and calculated torques have similar frequencies, but are related through a $\sim\pi/2$ phase shift.}
    \label{fig:torques_on_ski}
\end{figure}

Figure \ref{fig:torques_on_ski} shows a comparison of the torques calculated using Eqn. \ref{eq:torque_eqn} and the smoothed angular velocities for one ski run. The zoomed inset shows that the overall shape and frequency of the torque and angular velocity curves are similar except for some scaling and a phase shift of approximately $\pi/2$ radians, suggesting a simple first order linear relationship between torque and angular velocity. Specifically, the $\pi/2$ shift suggests that the torque is directly related to the derivative of angular velocity, something we explore in more detail in the following paragraphs. 

Although torque is indirectly available through Eqn. \ref{eq:torque_eqn}, these calculations require numerical differentiation and multiplication terms, which can both amplify sensor noise. To avoid using the highly processed torque signal, we take a closer look at the data and torque equation to find opportunities for simplifying the calculations. We begin by examining the terms of Eqn. \ref{eq:torque_eqn}. The IMU was positioned close to the center of mass, allowing us to simplify Eqn. \ref{eq:torque_eqn} to: 

\begin{equation}
  \label{eq:torque_eqn_simple_1}
  \prescript{N}{}{\Vec{M}}^{B/B_p} = \Vec{\Vec{I}}^{B/B_p} \cdot \prescript{N}{}{\Vec{\alpha}}^B + \prescript{N}{}{\Vec{\omega}}^B \times (\Vec{\Vec{I}}^{B/B_p} \cdot \prescript{N}{}{\Vec{\omega}}^B).  
\end{equation}
To gain better intuition for Eqn. \ref{eq:torque_eqn_simple_1} we explicitly write out the moment about one axis:

\begin{equation}
\label{torque_x_simple}
\footnotesize
\begin{centering}
\begin{multlined}
    M_{x} = I_{xx}*\Dot{\omega}_x+
    \underbrace{\omega_x\omega_y(I_{zz}-I_{yy})}_\text{Multiplication of angular velocities} + \\
    \underbrace{I_{yz}(\omega^{2}_y-\omega^{2}_z)-\omega_x\omega_xI_{xy}+I_{xy}\Dot{\omega}_y.}_\text{Terms associated with products of inertia}
\end{multlined}
\end{centering}
\end{equation}

Here we see that several terms include products of inertia (e.g. $I_{yz}, I_{xy}$). From Table \ref{tab:inertia_dyadic} we can see that these products are all much smaller than terms like $I_{xx}$, and will therefore only make a minor contribution to the torque, allowing us to ignore these terms. Furthermore, from Fig. \ref{fig:torques_on_ski} we can see that $\omega_y$ is consistently much smaller than $\omega_x$ and $\omega_z$, suggesting that any terms that contain a product with $\omega_y$ will be small. Together, these observations suggest that, for our system, $M_x$ is primarily driven by the angular acceleration about the $x$ direction, $\dot{\omega}_x$. Similar arguments can be constructed to justify that $M_y$ is largely driven by $\dot{\omega}_y$, and $M_z$ is largely driven by $\dot{\omega}_z$. 

\begin{figure}[H]
    \centering
    \includegraphics[scale=1]{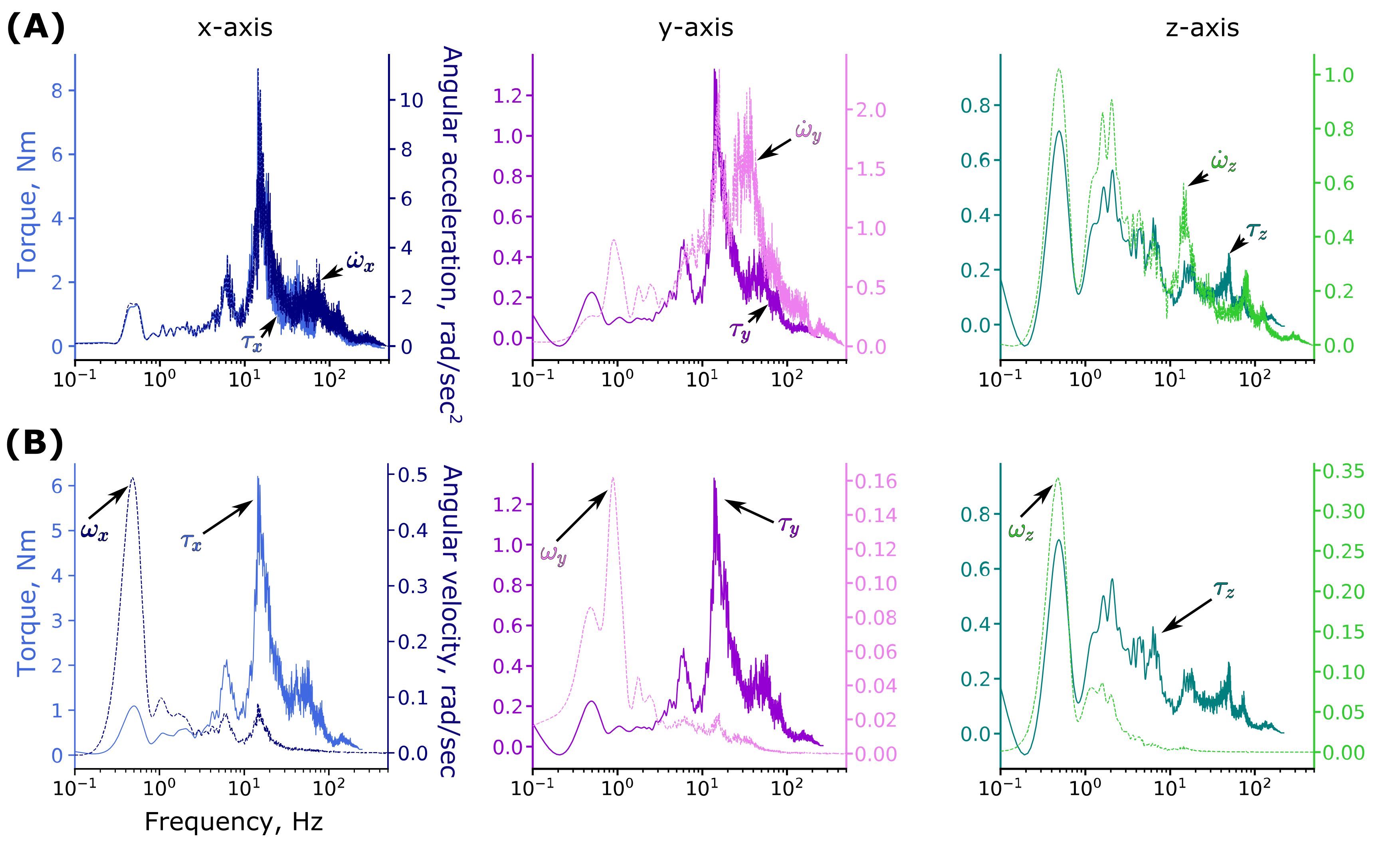}
    \caption{\footnotesize (A) The DFT of torque about each axis (e.g. $\tau_{x}$) is generally well approximated by a scaled DFT of angular acceleration about the same axis (e.g. $\dot{\omega}_x$), though some disagreement can be seen for the y-axis. (B) The relationship between torque and angular velocity ($\omega$) is similar, with $\omega$ scaled by the discrete frequency array $\Omega$. This frequency dependent scaling of $\omega$ means that at low frequencies the magnitude of $\omega$ is comparatively larger than $\tau$, whereas at high frequencies the magnitude of $\omega$ is attenuated relative to $\tau$. The DFT's shown here correspond to the timeseries data in Fig. \ref{fig:torques_on_ski}.}
    \label{fig:dft_tau_omega}
\end{figure}

Our analysis thus far allows us to approximate the torque inputs using much simpler angular accelerations. Thus, to construct our frequency domain models we can justify calculating $DFT(\dot{\omega})$ instead of $DFT(\tau)$. Figure \ref{fig:dft_tau_omega}A shows a comparison of the $DFT$ of $\tau$ and $\dot{\omega}$, confirming that for our system--and in the context of our frequency domain analysis--it is reasonable to approximate the torques using angular accelerations.  

Estimating the angular acceleration given noisy angular velocity measurements from the IMU, however, still amplifies sensor noise. To address this issue, we note that the derivative can be calculated in the Fourier domain as $DFT(\frac{d}{dt}f(t)) = j\Omega DFT(f(t))$, where $j\Omega$ is the discrete frequency array corresponding to the $DFT$. Since we only analyze the real component going forward, we drop the $j$, allowing us to use $\Omega DFT(\omega)$ instead of $DFT(\dot{\omega})$. This multiplication still increases the high frequency noise. Since $\Omega$ remains consistent across all of our experiments, and our goal is to compare the data-driven models associated with different 10 second segments, we can justify dropping the multiplication by $\Omega$ from our models for the purposes of using the models solely for classification. Figure \ref{fig:dft_tau_omega}B shows a comparison of $DFT(\omega)$ and $DFT(\tau)$.

In summary, we can build approximate frequency-scaled linear models of the ski-snow interactions associated with each strain gauge (e.g. $SG_{F||})$ to each axis of actuation by calculating the following DFT ratios:

\begin{equation}
\label{eq:dft_strain_omega}
\begin{split}
\text{ski-snow interactions for } SG_{F||} = \\
    \bigg[ \frac{DFT(SG_{F||})}{DFT(\omega_x)}; \frac{DFT(SG_{F||})}{DFT(\omega_y)}; \frac{DFT(SG_{F||})}{DFT(\omega_z)} \bigg].
\end{split}
\end{equation}
Although our analysis focuses on using angular velocity as the input from this point forward, we do compare our final classification results to those achieved by using either the full torque calculations or angular accelerations as the inputs.

\subsubsection{Calculating DFT ratios}
\label{methods_dft_ratio}
To model the ski-snow system we numerically calculated the real component of the ratio of the Discrete Fourier Transform (DFT) of each strain gauge and the angular velocity around each axis. The DFTs were calculated using the SciPy library \citep{2020SciPy-NMeth}. Solving for the ratio can be na\"{\i}vely done by solving the following,

\begin{equation}
    \mathcal{A} \mathcal{X} - \mathcal{B} = 0,
    \label{eq:cvx_py_simple}
\end{equation}
where $\mathcal{A}$ is the DFT of the angular velocities, $\mathcal{B}$ is the DFT of the output strain, and $\mathcal{X}$ is the DFT ratio. This ratio is essentially the magnitude portion of the Bode plot relating the strain and angular velocity. We use the term DFT ratio to (a) emphasize that we only focus on the magnitude portion (due to constraints imposed by sensor noise), and (b) as described in the previous section we use the angular velocity instead of the true input of torque. 

\begin{figure}[H]
    \centering
    \includegraphics[scale=1.25]{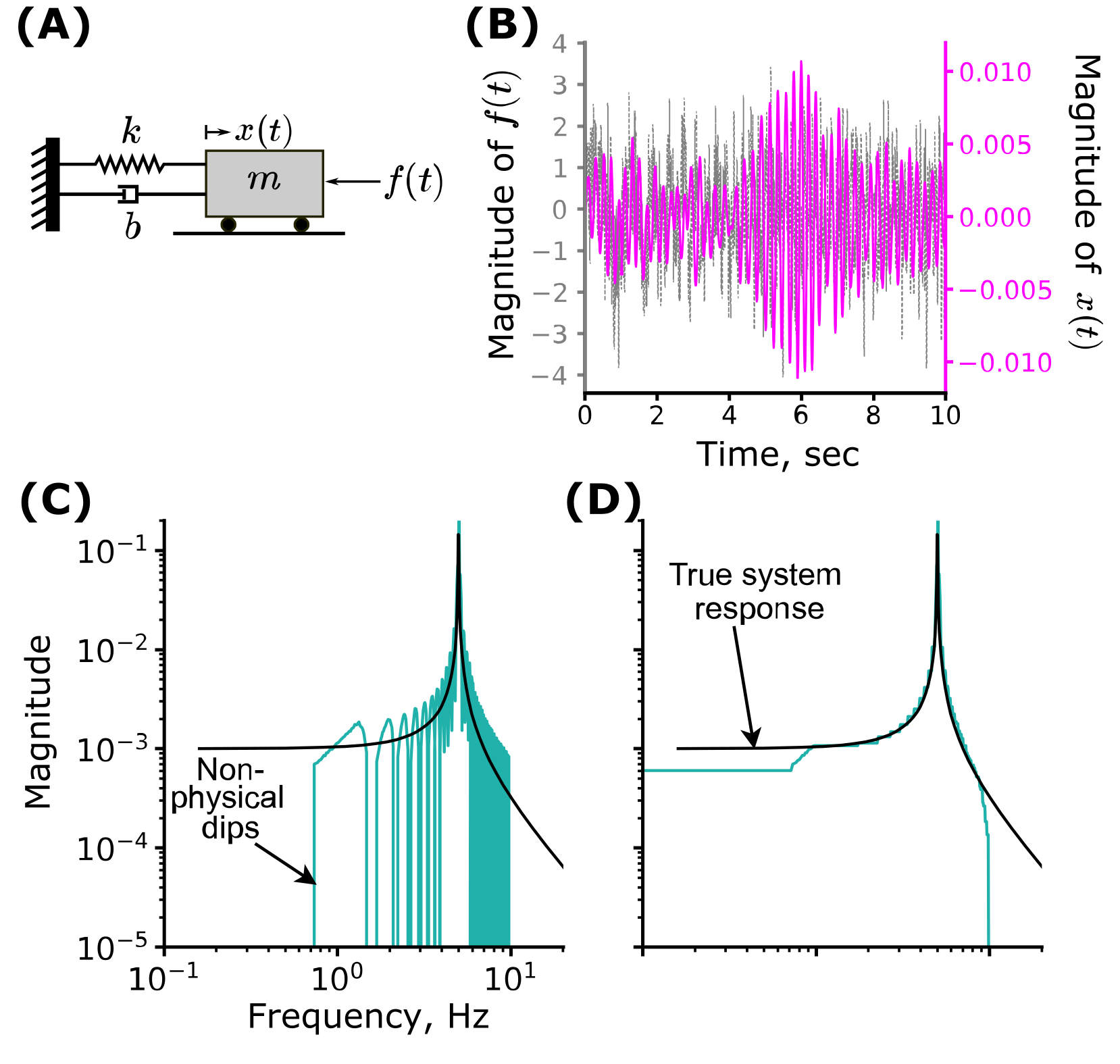}
    \caption{\footnotesize Non-physical dips in the DFT ratio are eradicated when penalizing the total variance. (A) Diagram of a mass spring damper system used to test our method, k=25 N/m, m=1 kg, b=1 Ns/m. (B) Time trace showing the forcing function $f(t)$ given by a noisy chirp signal containing frequencies from 0.1-20Hz. (C) Teal trace shows the DFT ratio calculated using the na\"{i}ve division approach (Eqn. \ref{eq:cvx_py_simple}). (D) Teal trace shows the DFT ratio calculated after adding a total variation regularization term (Eqn. \ref{eq:loss_function}). The peak seen in both plots align at the resonant frequency of the system. In both panels the black trace shows the true no noise magnitude portion of the Bode plot calculated using classic systems engineering techniques.} 
    \label{fig:simulate_tv}
\end{figure}

Calculating $\mathcal{X}$ through na\"{\i}ve element-wise division on real data with noise can lead to non-physical results which appear as jagged dips approaching zero caused by a small number being divided by a large number. To illustrate this potential error we ran a simple mass spring damper simulation with an input force given by a noisy chirp signal (Fig. \ref{fig:simulate_tv}A-B). The na\"{\i}vely calculated DFT ratio of the output (mass position) and  input (force) for this system is shown in Fig. \ref{fig:simulate_tv}B. 

The non-physical jumps would be difficult to eliminate simply by smoothing the DFT ratio, so we took a different approach: we added a total variation regularization term to Eqn. \ref{eq:cvx_py_simple}. Total variation regularization helps to minimize the difference between neighboring points \citep{agrawal2018rewriting}, penalizing the non-physical dips. We can write the regularized version of Eqn. \ref{eq:cvx_py_simple} using a loss function:

\begin{equation}
    \mathcal{L} = ||\mathcal{(AX - B)}^{2} + \gamma TV\mathcal{(X) }||_1,
    \label{eq:loss_function}
\end{equation}
where $\mathcal{L}$ is the loss function, $TV$ is the total variance, and $\gamma$ is a hyperparameter that determines how strong of a regularization to apply. To solve for $\mathcal{(X)}$, we use a convex optimization package in python (CVXPY \citep{diamond2016cvxpy}), with the proprietary solver (MOSEK \citep{mosek}). Applying this approach (with $\gamma=0.8$) on our simulation yields a DFT ratio that is much smoother and more representative of the real DFT ratio Fig. \ref{fig:simulate_tv}B. 

\subsection{Data analysis and classification}

Using the approach described above for calculating DFT ratios of each strain gauge relative to each of the three angular velocity components, we can construct a collection of three data-driven models for each strain gauge (e.g. Eqn. \ref{eq:dft_strain_omega}), for each 10 second segment from each ski run. In this section we describe our approach to training a classifier to assign qualitative labels to each model collection. We begin with a discussion of the raw data and example DFT calculations, followed by a description of a model reduction approach we use to simplify the inputs to our classifier, and finally the classifier itself.

\subsubsection{Raw timeseries data}

\begin{figure}[H]
    \centering
    \includegraphics[scale=0.9]{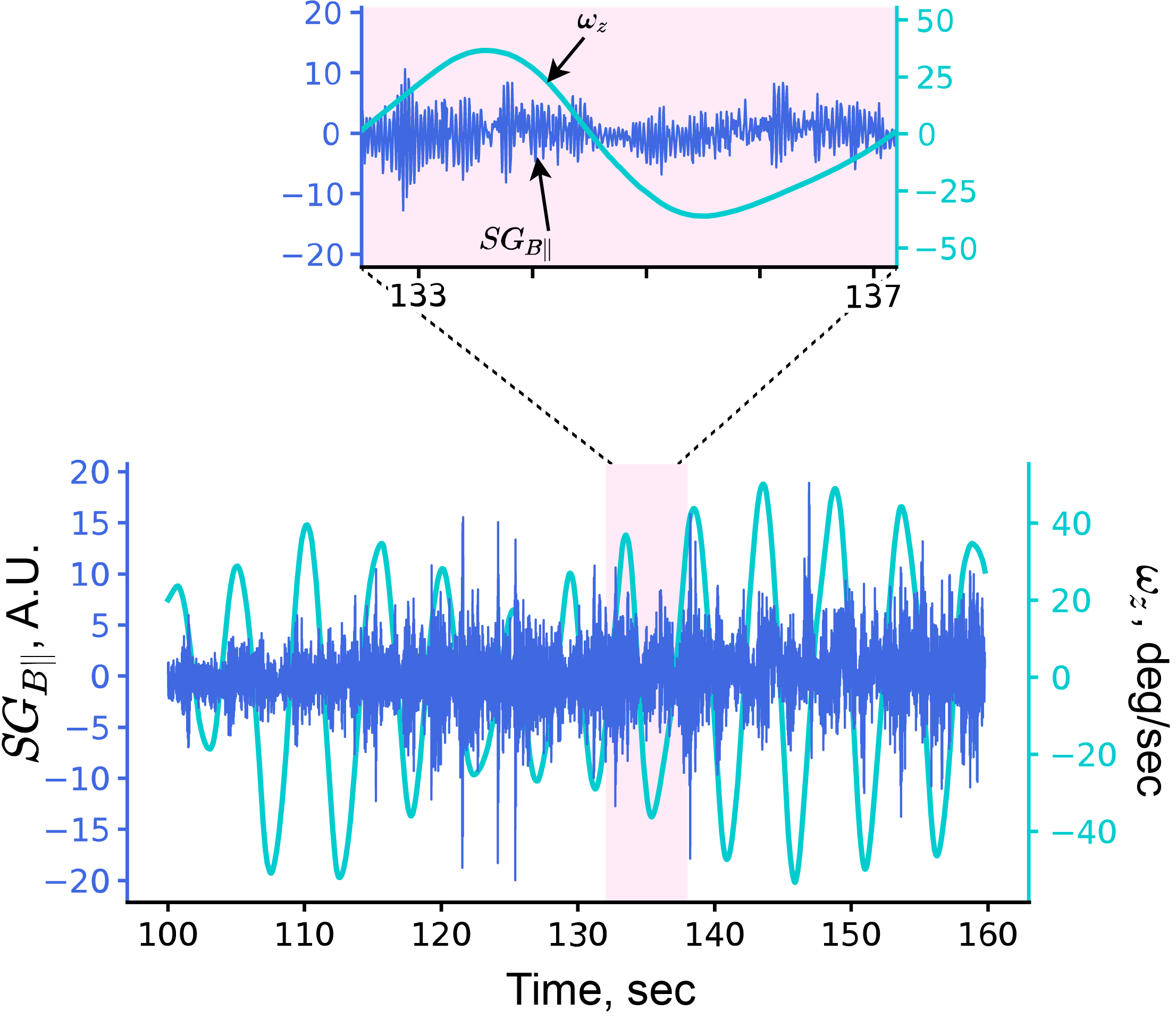}
    \caption{\footnotesize Data from a full ski run on groomed snow with big and slow turns. Smoothed angular velocity about the z-axis (cyan) and strain from the back parallel strain gauge (blue) are shown.}
    \label{fig:full_run_2turn}
\end{figure}

To generate spatially dense classifications of snow type we analyzed our data and trained our classifier using data segments with at least one consecutive pair of left/right or right/left turns. Such segmentation ensures that all bending modes of the ski are included. Figure \ref{fig:full_run_2turn} shows example raw strain data from the SG$_{B\parallel}$ in blue and the smoothed z-axis angular velocity in cyan from one ski run on groomed snow. Using the smoothed z-axis angular velocity trace, we determined that for ski runs with big and slow turns, trajectory snippets needed to be 10 seconds long to always contain a pair of left and right turns. For consistency across ski runs, we used 10 second segments for both the slow and fast turn trajectories. The strain signal shows highly varied frequency content including both high and low frequencies. 

\begin{figure}[H]
    \centering
    \includegraphics[width=0.95\textwidth]{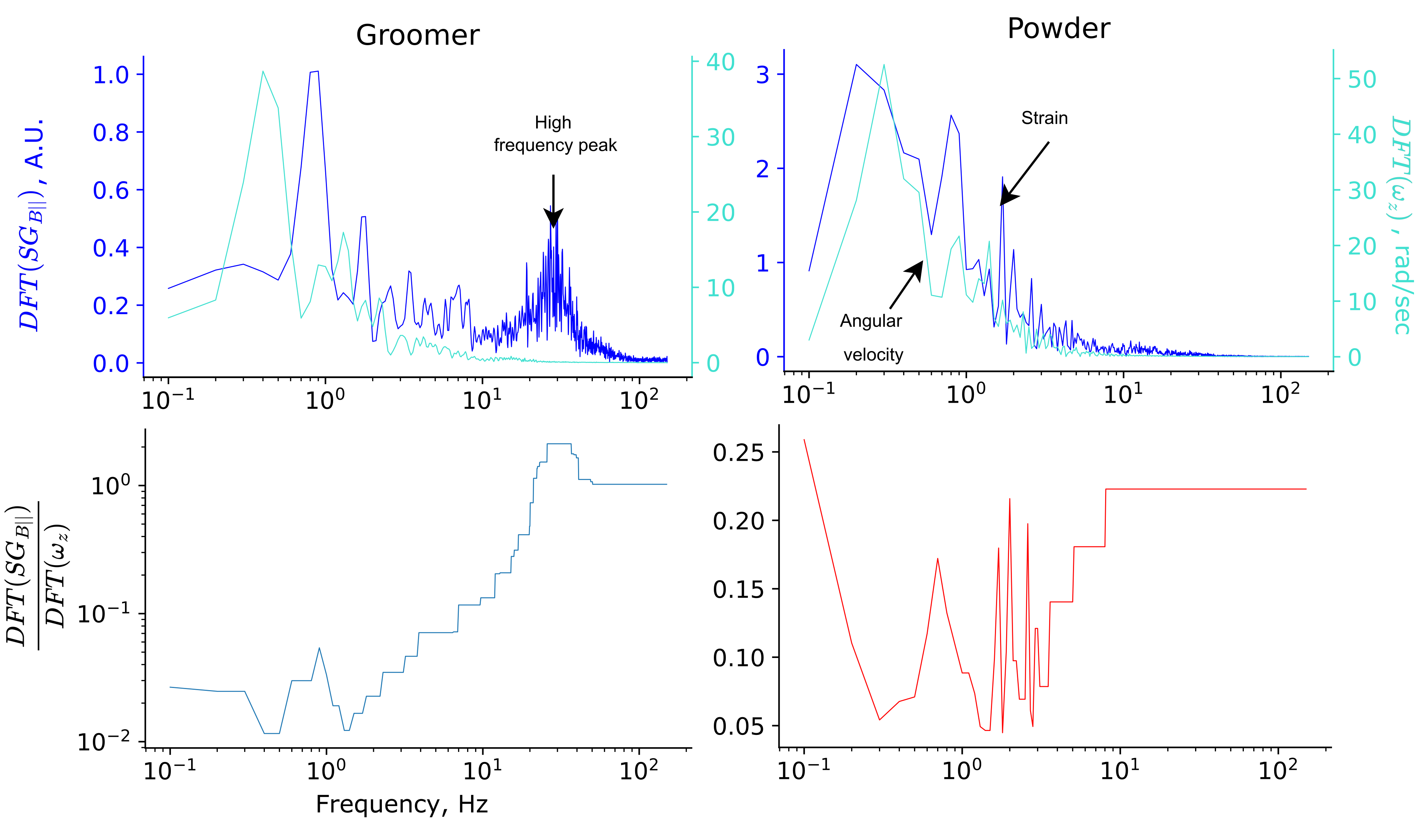}
    \caption{\footnotesize Comparison of strain, angular velocity, and DFT ratios for ski runs on groomer and powder snow. The groomer DFT has a high frequency spike and the powder does not.}
    \label{fig:dft_groomer_vs_powder}
\end{figure}

\subsubsection{Frequency content and DFT ratios}

The first step in our modeling and classification approach is to calculate the DFT of the strain signals and angular velocities. This step also helps to provide intuition into the differences that are found when comparing data from different snow types. Figure \ref{fig:dft_groomer_vs_powder} shows the DFT of the strain and z-axis angular velocity for one 10 second segment on both groomed and powder runs. 

A high frequency spike is apparent in the groomer data, which does not exist in the powder data (Fig. \ref{fig:dft_groomer_vs_powder}). This observation is consistent with the experience of feeling high frequency vibrations when skiing on harder pack snow compared to soft powder. The low frequency magnitude spike in both the angular velocity and strain follows our intuition that the skier imparts some signal onto the ski. Next we calculate the DFT ratio between the strain measurements and angular velocities using the approach described in Section \ref{methods_dft_ratio} with $\gamma= 0.8$ (Fig. \ref{fig:dft_groomer_vs_powder}). We manually selected $\gamma$ to reduce the non-physical dips in our data without sacrificing the overall shape of the DFT ratio. The DFT ratio largely eliminates the skier's input (as measured by the angular velocity) and allows us to focus on the ski-snow interaction.

\begin{figure}[H]
    \centering
    \footnotesize
    \includegraphics[scale=0.9]{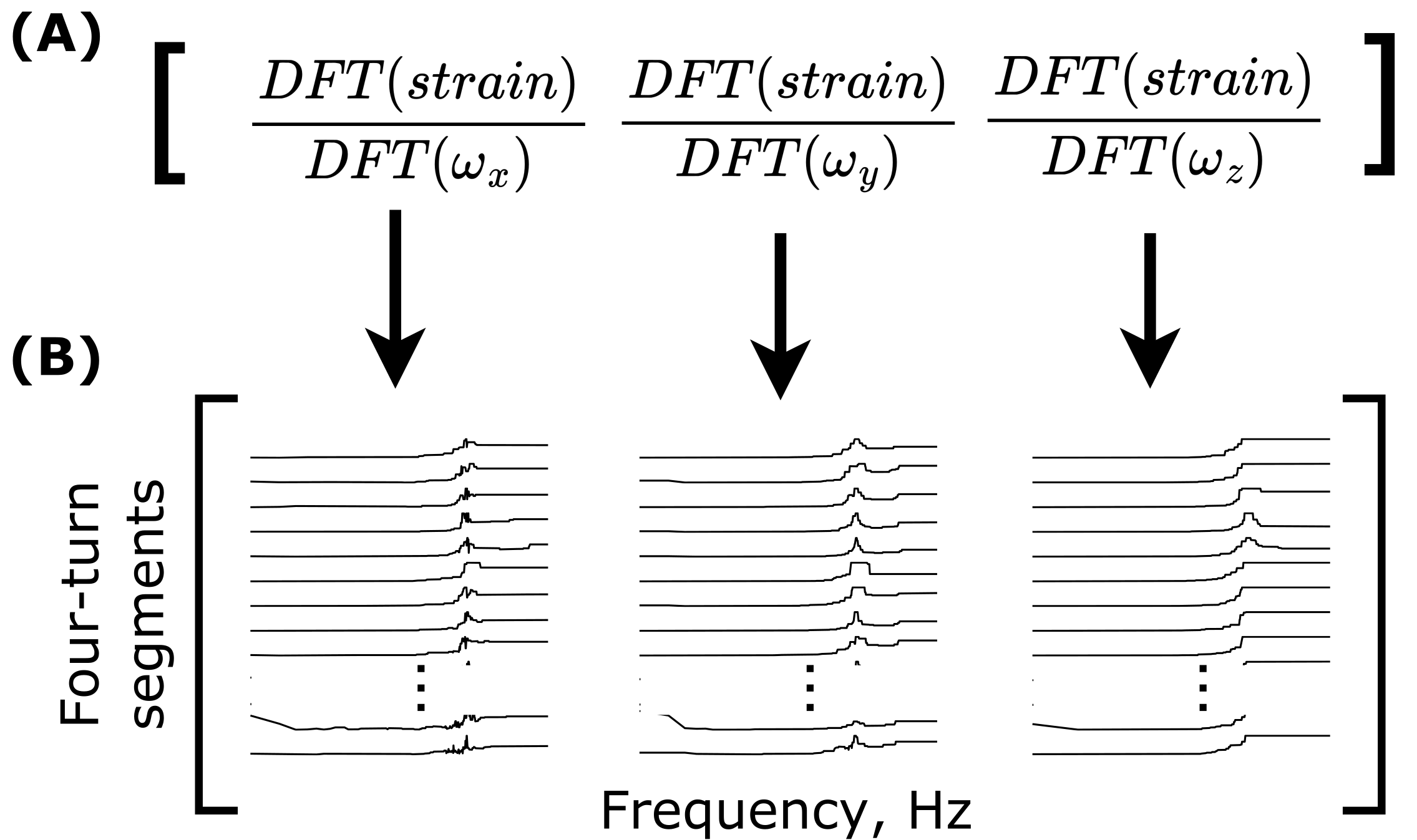}
    \caption{\footnotesize (A) DFT ratios are calculated for each angular velocity individually for each strain gauge and stacked horizontally in (B) to create the DFT ratio matrix.}
    \label{fig:bode_flow}
\end{figure}

To build our snow type classifier we first aggregate the data from an individual strain gauge for all of the 10 second segments from all ski runs into a single matrix. Each row of the matrix contains data from one 10 second segment, and consists of the three DFT ratios between the strain gauge and each angular velocity component concatenated together (Fig. \ref{fig:bode_flow}). The result is a 65x4497 matrix, where 65 is the total number of 10 second segments for all runs and 4497 is the number of DFT ratio data points. To help our classifier focus on the most important features of the data we used a Singular Value Decomposition to reduce the dimensionality of the DFT ratio matrix, as described in the following section.

\begin{figure}[H]
    \includegraphics[scale=0.6]{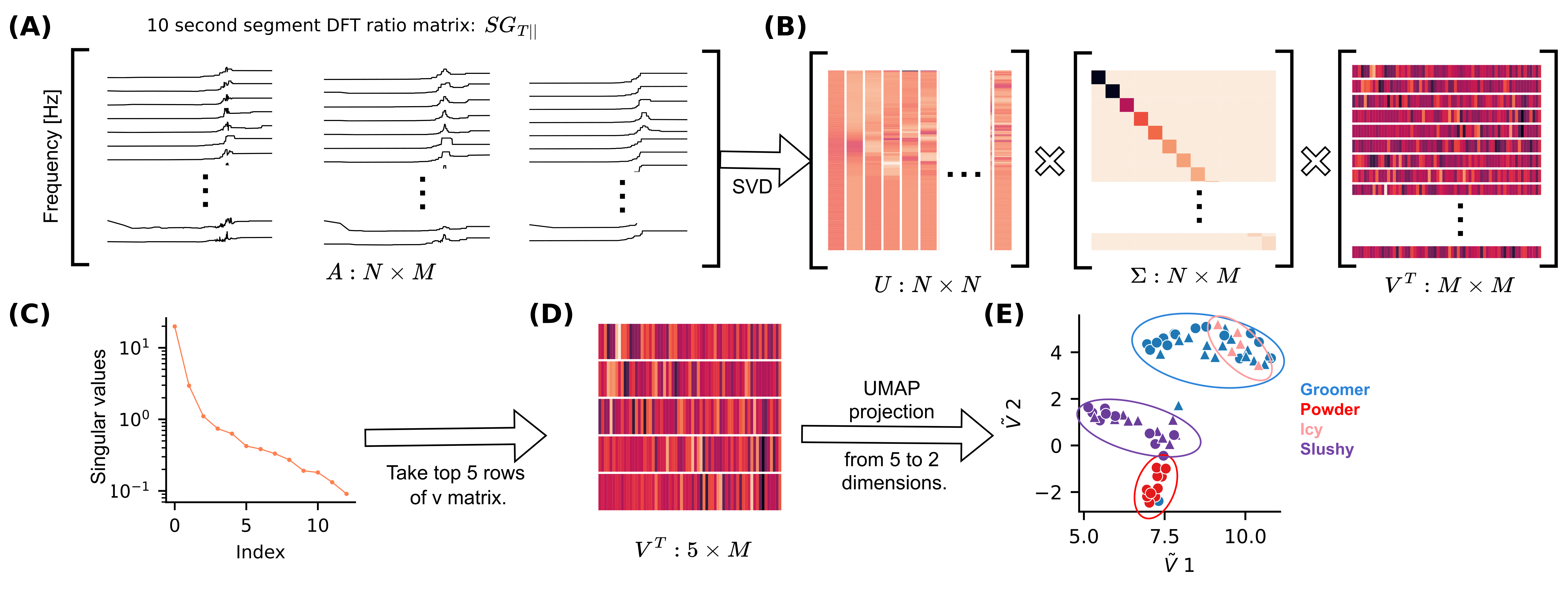}
    \centering
    \caption{\footnotesize Overview of our analysis pipeline. (A) Horizontally stacked DFT ratios in the frequency domain. (B) SVD decomposition, where U is the new set of bases that the data exists in, S is a diagonal matrix of singular values, and V is a matrix of features in the SVD space. (C) Logarithmically scaled plot of the singular values showing that they drop in value very quickly. (D) Top five rows of the V matrix corresponding to the five largest singular values. (E) UMAP projection from the 5D space to 2D space.}
    \label{fig:svd_analysis}
\end{figure}

\subsubsection{Singular Value Decomposition and Projection}

To develop our classifier we first applied a Singular Value Decomposition (SVD) to find the dominant coherent structures in our DFT ratio matrix. In our case the coherent structures correspond to dominant shapes of the DFT ratios. A shape with a high frequency peak will most likely be associated with harder snow, whereas a DFT ratio with a lower frequency peak or no peak will correlate to softer snow.

The SVD performs an eigen decomposition of the original non-symmetric matrix $A$ (our DFT ratio matrix, Fig. \ref{fig:svd_analysis}A), resulting in three new matrices $U, \Sigma$ and $V$ (Fig. \ref{fig:svd_analysis}B) that are related according to the following expression \citep{baker2005singular}:

\begin{equation}
    A = U \Sigma V^T. \\
    \label{SVD_eqn}
\end{equation}The matrix $U$ has shape $N \times M$ with columns that describe the left singular vectors of $AA^{T}$, and represents the transformed set of bases where the coherent structures exist. $V^T$ is an $M \times M$ matrix whose rows are the right singular values of $A^{T}A$, corresponding to the weights necessary to reconstruct $A$ given the bases from the columns of $U$. In our case, each column of $V^T$ corresponds to one of the 10 second segments. $\Sigma$ is an $M \times M$ matrix consisting of the singular values of $A$ on the diagonals, which give the ranked importance of the rows of $V^T$. If all of the rows from $V^T$ are considered, it is possible to perfectly reconstruct $A$ using Eqn. \ref{SVD_eqn}. However, often most of the coherent structures in $A$ can be recovered by choosing just the first few rows of $V^T$, making it possible to reduce the dimensionality of the original data matrix, which reduces the number of features our classifier has access to leading to less over-fitting. Based on the rapid drop off of the singular values (Fig. \ref{fig:svd_analysis}C), we initially chose to focus on just the first five rows of the $V^T$ matrix (Fig. \ref{fig:svd_analysis}D). 

The rows of $V^T$ can be thought of as features that describe where the original data exist in a new reduced order basis. Five features are still too many to visualize in a two-dimensional figure. Therefore, to visualize how our DFT ratios are distributed in the reduced order space we projected the five features into a two-dimensional space using Uniform Manifold Approximation and Projection (UMAP) (Fig. \ref{fig:svd_analysis}E). The UMAP makes it possible to visualize the higher-dimensional space in two-dimensions while preserving the Euclidean distance \citep{mcinnes2018umap}. We used the UMAP implementation given in \citep{sainburg2021parametric}.

To summarize, for an individual strain gauge we calculate the DFT ratio matrix that includes data from each 10 second segment across all our ski runs, perform the SVD to map the data into a new orthonormal basis, choose the top five rows of $V^T$, and plot their two-dimensional projection using UMAP. The data from each individual 10 second segment results in a single point in the resulting scatter plot (Fig. \ref{fig:svd_analysis}E). Note how the data from the powder and slushy ski runs cluster separately from one another, and from the hard packed runs done on groomer and icy snow. Meanwhile, the icy and groomer data cluster together, as expected, since these snow types are in fact quite similar to one another. The clearly distinguishable clusters are what make possible to train a classifier to automatically assign qualitative snow type labels to each 10 second segment. In the results section we explore how classification can be improved by combining data from multiple strain gauges, which helps to separate the icy and groomer data from one another.

\subsubsection{Classification}

Although visually we can see that there are three distinct groups of data that have formed according to snow type in Figure \ref{fig:svd_analysis}E, we wanted to automate the classification process and quantify the separation of the clusters in a statistically rigorous manner by training a classifier through supervised learning. We trained a Gaussian Na\"{\i}ve Bayes classifier with 70\%  of the data, selected randomly, from the top five rows of $V^T$, and then tested our classifier's performance using the remaining 30\% using the Scikit-learn library \citep{scikit-learn}. The Gaussian Na\"{\i}ve Bayes classifier provided a more parsimonious decision boundary compared to other methods we explored such as the Support Vector Machine. 

We trained the Gaussian Na\"{\i}ve Bayes classifier using different combinations of strain gauges. This was done by concatenating the DFT ratio matrices for different strain gauges together prior to performing the SVD. The rows of the $V^T$ matrix from the SVD operation were used as features. Two feature decisions are made here. One is the combination of strain gauges, and the other is the quantity of features used to train the classifier. We analyze the impact of both decisions in the Results section. 

\section{Results}

\subsection{Skier influence on snow clustering}

\begin{figure}[H]
    \centering
    \includegraphics[scale=0.9]{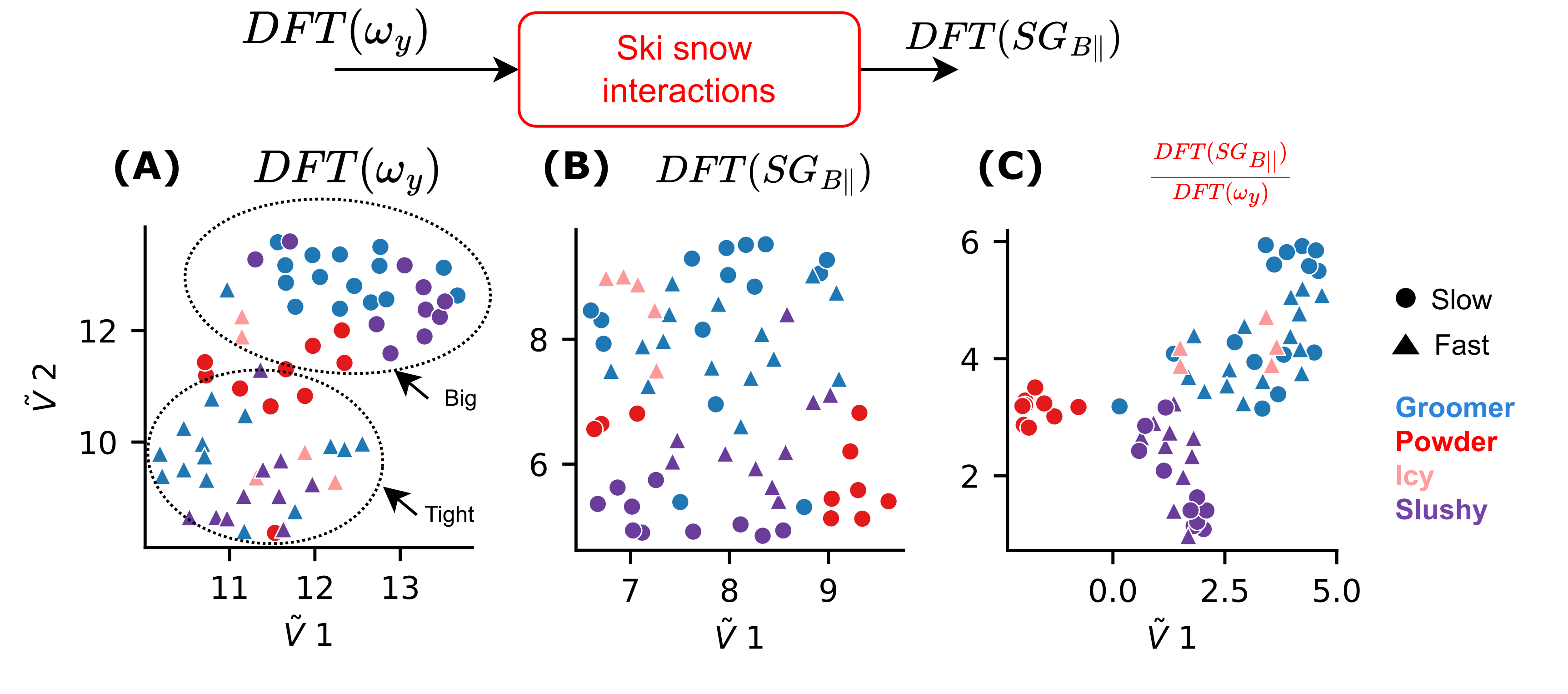}
    \caption{\footnotesize  Controlling for skiing style is crucial for achieving a clear clustering of data according to snow type. (A) Scatter plot showing separation of angular velocities based on skiing style. Triangles represent tight turns and dots represent big turns. (B) When just the strain signal is used, the skiing styles mix and snow based clusters begin to form, but are not clearly separated. (C) Using the DFT ratio of angular velocity and strain, snow based clusters are much more distinct.}
    \label{fig:scatter_plots_ang_sg_bode}
\end{figure}

The time-varying curvature of an alpine ski is a function of both skier-applied torques, and ski-snow interactions (recall Fig. \ref{fig:block_diagram_complete}A). In Section \ref{Modeling} we describe our modeling framework for removing the influence of the skier from our subsequent analysis and classification. To understand how important removing the skier influence is for accurate clustering based on snow type, we first applied the approach outlined in Fig. \ref{fig:svd_analysis} to analyze the DFTs of either the angular velocity or strain without first calculating their ratio. As expected, when we analyze the DFT of the angular velocity for each 10 second segment, the data self-organize according to skiing style in the reduced order SVD space, but not snow type (Fig. \ref{fig:scatter_plots_ang_sg_bode}A), confirming that the angular velocity signal is strongly correlated with the skiing style. When we analyze the DFT of a strain gauge ($SG_{B||}$), on the other hand, the data do not cluster well according to either the skiing style, or snow type (Fig. \ref{fig:scatter_plots_ang_sg_bode}B), confirming that the strain signal is not dominated by one or the other. Finally, when we analyze the DFT ratio of a strain gauge and angular velocity (in this case $DFT(SG_{B||})/DFT(\omega_y)$), we see that the data cluster well according to snow type, and the effect of skiing style is effectively removed (Fig. \ref{fig:scatter_plots_ang_sg_bode}A). 

\begin{figure}[H]
    \centering
    \includegraphics{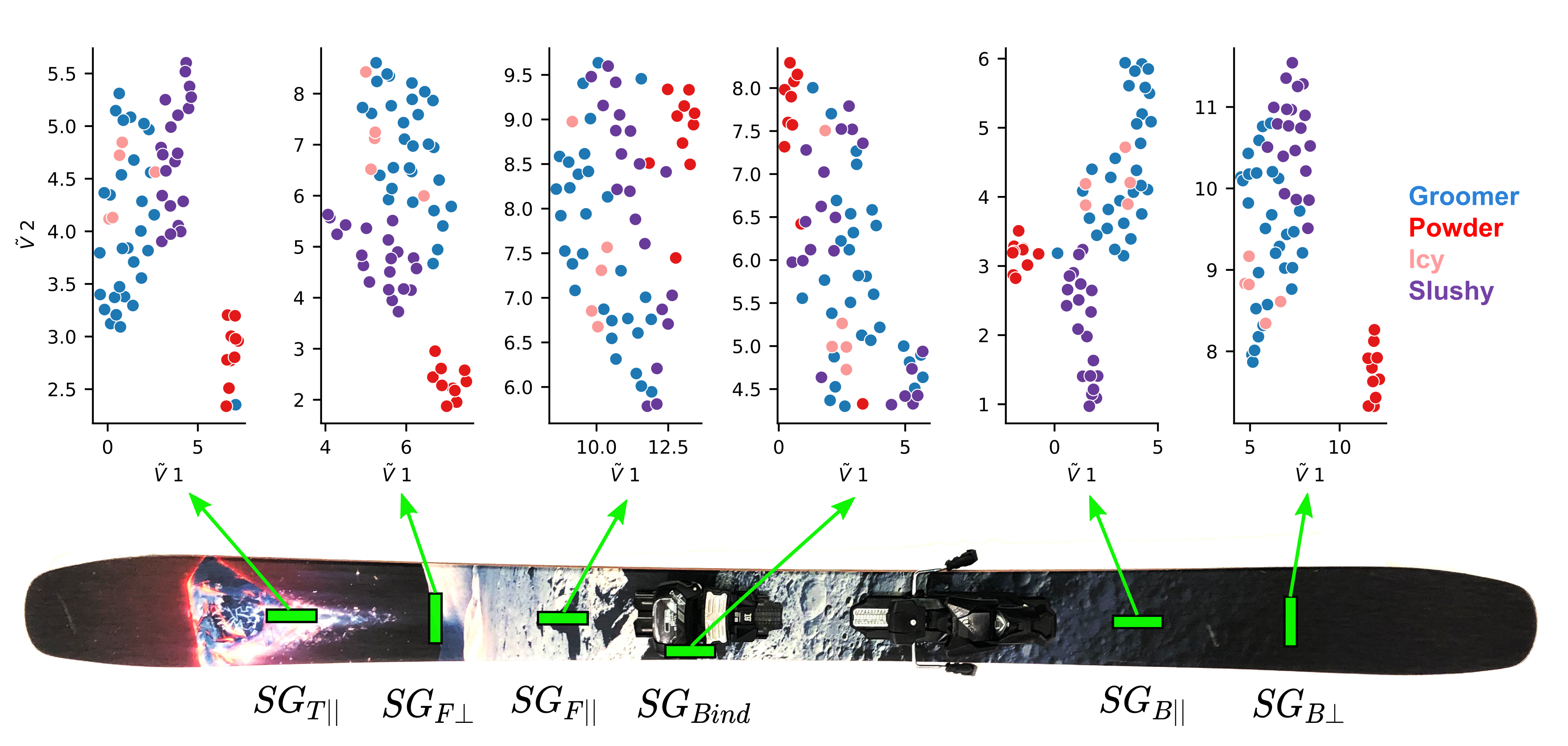}
    \caption{\footnotesize Clusters of ``V" rows of SVD projected into a two-dimensional UMAP space for each strain gauge reveal location specific behavior. The strain gauges closer to the binding cluster poorly compared to the strain gauges closer to the tip.}
    \label{fig:all_sgs_clusters}
\end{figure}

\subsection{Analysis of Individual Strain Gauges}

When designing our hardware and sensor arrangements, we hypothesized that different locations on the ski, and different orientations of strain sensors, might provide different types of information. To test this theory, we individually analyzed the DFT ratios for each strain gauge (with respect to all three axes of angular velocity) using the analysis approach outlined in Figure \ref{fig:svd_analysis}; the resulting cluster plots are shown in Figure \ref{fig:all_sgs_clusters}. The data from $SG_{Bind}$ and $SG_{F||}$ do not result in clear clusters according to snow type, except for the powder snow (red) which is grouped together but sill near the rest of the data points. Data from both strain gauges on the back of the ski, $SG_{B||}$ and $SG_{B\perp}$, do show clear clusters according to snow type. $SG_{B||}$ shows a clearer separation between the slushy and hard packed snow types (groomer, icy), whereas $SG_{B\bot}$ results in a larger separation between the powder snow and the rest. The data from the front of the ski, $SG_{T||}$ and $SG_{F\perp}$, also display distinct snow based clusters, with the data from powder snow separating especially clearly from the rest. Overall, data from strain gauges closer to the tip and tail form more distinct clusters according to snow type compared to data from strain gauges near the center of the ski. 

\begin{figure}[H]
    \centering
    \includegraphics[scale=1]{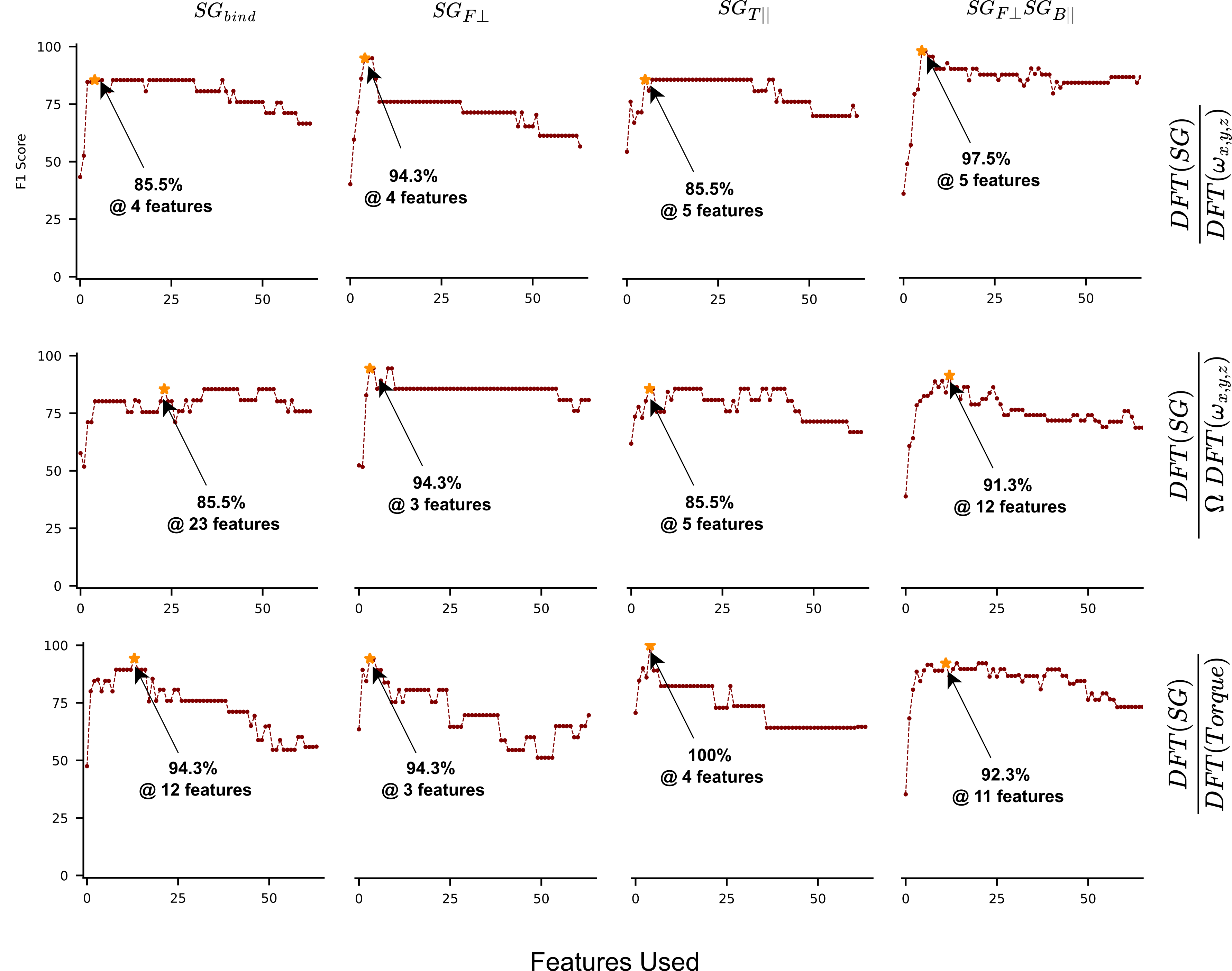}
    \caption{\footnotesize $F_1$ score of the Na\"{\i}ve Bayes classifier for individual strain gauges, and one combination, for the three different $DFT$ ratio options we consider. }
    \label{fig:classifier_plots}
\end{figure}

\subsection{Na\"{\i}ve Bayes Classification}

To automatically assign a qualitative snow type label to each 10 second segment, and to provide a numerical metric for the separation of the snow based clusters seen in Fig. \ref{fig:all_sgs_clusters}, we implemented a Na\"{\i}ve Bayes classifier. For this analysis, rather than solely focus on the first five features (the rows of the $V^T$ matrix) as we have done thus far, we trained and tested the classifier using anywhere from 1-80 features to find the quantity of features that gave the highest classification percentage. Using too many features can lead to a decrease in classifier performance, referred to as \textit{The Curse of Dimensionality} \citep{1054102}.

\begin{figure}[H]
    \centering
    \includegraphics{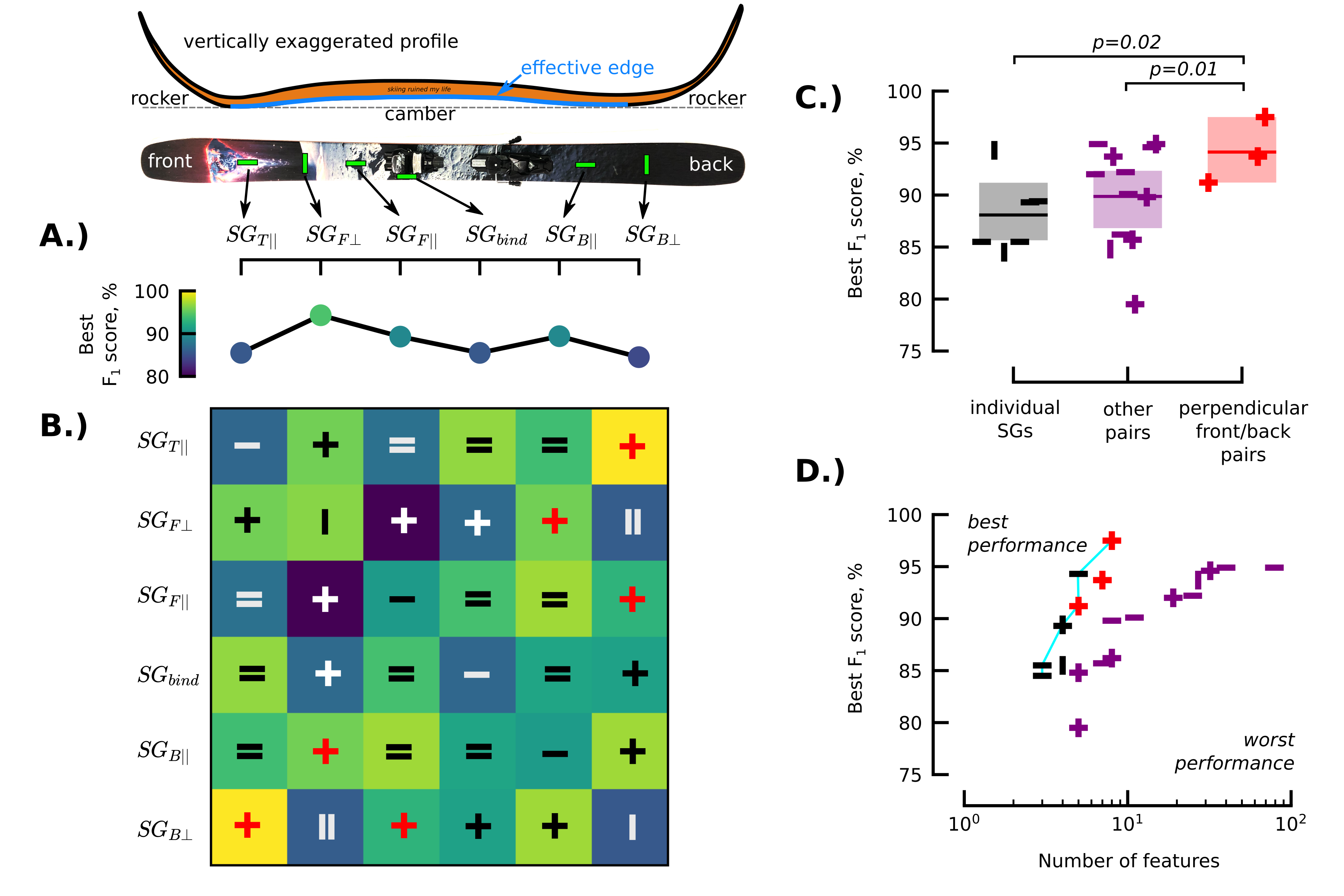}
    \caption{\footnotesize Strain gauge combinations with perpendicular orientations placed on opposite ends of the ski provide the best performance compared to either individual strain gauges and other pairs. A) Best $F_1$ score for individual strain gauges as a function of location and orientation on the ski. B) Best $F_1$ score for all possible pairwise combinations of strain gauges. Color indicates $F_1$ score using the colormap from panel A. Symbols indicate whether pairs consist of two strain gauges that are parallel to the ski ($=$), perpendicular to the ski ($||$), or strain gauges that are perpendicular to each other ($+$). C) Distribution of the best $F_1$ scores split into three groups: individual strain gauges (black), pairs of strain gauges that are both perpendicular to each other, and placed on opposite ends of the ski (red), and all other combinations (purple). The perpendicular front/back group results in a statistically significant improvement in $F_1$ score over individual strain gauges (p=0.05, resampling test with 10,000 iterations). Shading indicates the 95\% confidence interval of the mean, which is indicated by the bar. D) Best $F_1$ score plotted as a function of the number of features required to achieve that $F_1$ score. Points towards the upper left represent ideal combinations where high $F_1$ scores can be achieved with a relatively low number of features. Cyan line indicates the pareto front that balances the $F_1$ score and number of features.}
    \label{fig:ski_summary}
\end{figure}

To assess the performance of our classifier, we calculated the $F_1$ score, which combines two metrics, precision and recall, to provide a single metric that takes into account both false positives and false negatives. We trained our classifier on 70\% of the data, and tested on 30\%. Figure \ref{fig:classifier_plots} shows how $F_1$ score varies with the number of features for a few individual strain gauges, and one pair. Figure \ref{fig:ski_summary} shows the $F_1$ score for each individual strain gauge, and all possible combinations. 

Data from individual strain gauges that were placed near the inflection points of the effective edge--just center of the points where the unweighted ski has the highest contact with the snow--resulted in the clearest separation of snow types (Fig. \ref{fig:all_sgs_clusters}) and the highest $F_1$ scores (Fig. \ref{fig:ski_summary}A). Data from strain gauges in the middle of the ski (at the top of the camber) resulted in the worst separation of snow types, and low $F_1$ scores. Tip and tail strain gauges placed right at the point where the unweighted skies touch the ground did separate powder from other snow types (Fig. \ref{fig:all_sgs_clusters}), but were less effective at separating slushy snow from groomer and icy snow, resulting in a lower overall $F_1$ score (Fig. \ref{fig:ski_summary}A). For all strain gauge locations, distinguishing icy from groomer snow was the most challenging classification task, which is unsurprising given the relative similarity in the skiing experience over these two types. Combining data from certain pairs of strain gauges improved $F_1$ performance (Fig. \ref{fig:ski_summary}B). Broadly, the inclusion of one strain gauge parallel to the ski, and one perpendicular, from opposite sides of the ski, resulted in the highest $F_1$ scores for the least number of features (Fig. \ref{fig:ski_summary}C-D). The improved performance from including two strain gauges in opposite orientations likely provides better measurement coverage of the bending and torsional modes, which prior studies have shown are coupled together \citep{gosselin2021effect}. 

Finally, we revisit our proposition that using angular velocity for our input, as a proxy for torque, offers some advantages by reducing the quantity of noise amplifying steps in the analysis. We repeated the DFT ratio calculations using either angular acceleration or torque, followed by our dimensionality reduction and classification approaches. Figure \ref{fig:classifier_plots} shows the results when using the angular acceleration as the input, i.e. using $\Omega DFT{\omega_{[x,y,z]}}$ in the denominator of the DFT ratio calculations, or our torque estimates. In many cases, these ratios require more features to achieve similar performance compared to the ratio that uses angular velocity in the denominator. For example, the combination of $SG_{F\perp}$ and $SG_{B||}$ achieves a performance of 97.5\% with just five features with angular velocity in the denominator, whereas using the acceleration or torque requires more than twice as many features to reach 91-92\%. Using the $DFT(\omega_{[x,y,z]})$ for the DFT ratio likely provides the best performance for the least number of features because it is the least processed signal.

\section{Discussion}

To classify snow into qualitatively different groupings given data from a sensor equipped ski, we developed a three-step algorithm. First, we found a regularized discrete fourier transform (DFT) ratio comparing the frequency content for six strain gauges and the three axes of angular velocity of the ski's centroid. This ratio serves as a linear data-driven model (e.g. a numerical description of the real component of a transfer function) that describes the dynamics of the ski-snow interaction. We created such a model for each 10 second segment from several ski runs on different snow types. Next, we reduced the dimensionality of the resulting models using a singular value decomposition to extract a sparse collection of features. Finally, we trained a Na\"{i}ve Bayes classifier on this sparse feature set to assign a qualitative snow-type label to each 10 second segment, and validated our results using a separate test set. By combining information from two strain gauges, our algorithm was able to correctly classify snow types with an $F_1$ score of 97\% using 5-9 features, depending on the choice of strain gauges. 

Unsurprisingly, the key feature distinguishing the hardest snows (groomer and icy snow) from powder snow was the presence of a $\sim$30 Hz resonant peak for the hard snows, and a larger amplitude at low frequencies for the powder snow (Fig. \ref{fig:dft_groomer_vs_powder}), indicating fewer vibrations and more sustained deflections of the ski. The resonant peak we found for our ski runs on hard packed snow is comparable to previous studies that found resonant frequencies of 20-70 Hz \citep{gosselin2021effect, foss2007reducing, glenne1999ski}. 

Despite the numerical challenges associated with calculating the DFT ratio of the strain gauge measurements and the angular velocities (Fig. \ref{fig:simulate_tv}), this step proved crucial for accurately separating snow types (Fig. \ref{fig:scatter_plots_ang_sg_bode}). Although the angular velocity does not directly correspond to the physical input to the ski, we found that in many cases it worked slightly better than more directly correlated quantities such as torque or acceleration \ref{fig:classifier_plots}. We suspect this is the case because of the noise amplification that results from numerically differentiating the noisy measurements from the IMU to estimate angular acceleration or torque. 

In future implementations, to reduce the quantity of strain gauges and bandwidth of information, we suggest restricting data collection to $\sim$150 Hz and choosing just two strain gauges. Based on our analysis, choosing one parallel and one perpendicular strain gauge placed on the front and back of the ski near the inflection points of the camber appears to be an optimal choice. A simplified hardware implementation would make it more feasible to implement our design on backcountry skis. This would make it feasible to collect a more diverse set of training data including more varied terrain, slope angles, and snow surfaces. We anticipate that properly characterizing more snow types may require a more advanced modeling step, such as a nonlinear model that assigns a different DFT ratio depending on the skier's velocity. Simplifying the hardware required to collect the data would also make our approach more amenable to developing a product that could be used in a citizen science project that would allow backcountry skiers to contribute to a database of qualitative snow characteristics to compliment other remote sensing methods. For this to work, it is likely that either additional modeling or some calibration step would be needed to account for variations in skier weights, styles, and skis. Finally, we envision a long-term application where skis with tunable stiffness (e.g. using active damping \citep{rothemann2010active}) could be continuously and automatically adjusted to optimize the skis performance in varied conditions.  

\section*{Acknowledgements}
We are grateful to Moment Skis for donating the pair of Wildcat Skis to use for this project. We also thank Ryan Tung, Ben Cellini, and Burak Boyacıoğlu for comments on the manuscript, and Anne Nolin for advice throughout the project.

\section*{Funding Support}
This research was partially supported by the National Science Foundation AI Institute in Dynamic Systems (2112085).

\section*{Data Availability}
Data and code will be made available upon formal publication.


\appendix


\bibliographystyle{elsarticle-harv} 
\bibliography{cas-refs}





\end{document}